\documentstyle[sprocl,epsf]{article}

\def\be{\begin{equation}}
\def\ee{\end{equation}}
\def\bea{\begin{eqnarray}}
\def\eea{\end{eqnarray}}

\begin{document}

\begin{flushright}
e-print JLAB-THY-97-29 \\
July 1997 \\
\end{flushright}

\vspace{0.5cm}

\title{QCD SUM RULES: FORM FACTORS AND WAVE FUNCTIONS}

\author{ A.V. RADYUSHKIN 
\footnote{Also Laboratory of Theoretical Physics, JINR, 141980 Dubna, Russia}
}

\address{Old Dominion University, Norfolk, VA 23529, USA;\\
Jefferson Lab, Newport News, VA 23606, USA}


\maketitle\abstracts{ The shape of hadronic 
distribution amplitudes (DAs) is a critical issue
for the perturbative QCD of hard exclusive processes.
Recent CLEO data  on  $\gamma \gamma^* \to \pi^0$ form factor
clearly favor a pion DA close to the asymptotic form. 
We argue that   QCD sum rules for the moments of the
 pion DA $\varphi_{\pi} (x)$ 
are unreliable, so that the humpy shape of $\varphi_{\pi} (x)$
obtained by Chernyak and Zhitnitsky   
is a result of model assumptions rather than 
an unambigous consequence of QCD sum rules.  
This conclusion is also supported by a direct 
QCD sum rule calculation 
of the $\gamma \gamma^* \to \pi^0$ form factor
which gives a result very close to the CLEO data.
}

\section{Introduction} 

In this talk, I  discuss some general 
features of QCD sum rule applications 
to hadronic wave functions and form factors 
using as    examples  the pion distribution 
amplitude $\varphi_{\pi} (x)$
and transition form factor for the process 
 $\gamma^* \gamma^* \to \pi^0$ in which 
two virtual photons  produce  a neutral pion.
This process 
provides an exceptional opportunity to test
QCD predictions for exclusive processes.
In the lowest order of
perturbative QCD, its asymptotic behaviour
is due to the subprocess
$\gamma^*(q_1) + \gamma^*(q_2) \to \bar q(\bar xp) + q (xp) $
with $x$ ($\bar x$) being  the fraction of the pion momentum $p$ carried
by the quark produced at the $q_1$ ($q_2)$ photon vertex.
The  relevant diagram resembles
the handbag diagram of 
DIS 
with 
 the pion distribution amplitude (DA) $\varphi_{\pi}(x)$
instead of parton densities.
The asymptotic PQCD prediction is given by  \cite{bl80}\ ($q_1^2 \equiv -q^2$,
 $q_2^2 \equiv -Q^2$, $\bar{x}=1-x$):
\begin{equation}
F_{\gamma^* \gamma^*  \pi^0 }^{\rm as}(q^2, Q^2) = \frac{4\pi}{3}
\int_0^1 {{\varphi_{\pi}(x)}\over{xQ^2+\bar x q^2}} \, dx
\stackrel{q^2=0}{\longrightarrow}
\frac{4\pi}{3}
\int_0^1 {{\varphi_{\pi}(x)}\over{xQ^2}} \, dx
\equiv \frac{4\pi f_{\pi}}{3Q^2} I  .
\label{eq:gg*pipqcd}
\end{equation}
Experimentally,
the most important situation is when one of the photons
is almost real $q^2 \approx 0$  \cite{CELLO,CLEOnew}.
In this  case,  necessary nonperturbative information
is accumulated in
the same integral $I$  (see Eq.(\ref{eq:gg*pipqcd}))
which  appears  in the one-gluon-exchange
diagram  for the  pion electromagnetic  form factor
 \cite{pl80,blpi79,cz82}.

The  value of $I$ is sensitive to the shape of the
pion DA $\varphi_{\pi}(x)$.  Using  the
asymptotic  \cite{pl80,blpi79} form
$
\varphi_{\pi}^{\rm as}(x) = 6 f_{\pi} x \bar x
$
 gives $I=3$ and $F_{\gamma \gamma^*  \pi^0 }^{\rm as}(Q^2) =
4 \pi f_{\pi}/Q^2 $. If one takes the
CZ form \cite{cz82}
$\varphi_{\pi}^{\rm CZ}(x) = 30 f_{\pi} x\bar{x}(1-2x)^2$,
then  $I=5$,
and this difference can be used for experimental
discrimination between the two forms. One-loop radiative QCD corrections
to Eq.(\ref{eq:gg*pipqcd}) are known  \cite{braaten,kmr,musar} and they are
under control.
Clearly, the asymptotic $1/Q^2$-behaviour
cannot be true in the  low-$Q^2$
region,  since  the $Q^2=0$ limit of
$F_{\gamma \gamma^*  \pi^0 }(Q^2)$
is known to be finite and
normalized by the $\pi^0 \to \gamma \gamma$ decay rate.
From the axial anomaly \cite{ABJ},
$
F_{\gamma \gamma^*  \pi^0 }(0) =1/ \pi f_{\pi} .
$
Brodsky and Lepage \cite{bl80} proposed 
a simple interpolation
$
\pi f_{\pi} F^{\rm LO}_{\gamma \gamma^*  \pi^0 }(Q^2) =
1/(1+ Q^2/4 \pi^2 f_{\pi}^2)
$ 
between
the $Q^2=0$ value $1/\pi f_{\pi} $ and the leading-twist PQCD 
behavior $4 \pi f_{\pi}/Q^2 $ 
with normalization corresponding to 
the asymptotic DA. Note that the  mass scale 
$s^{\pi}_o \equiv 4 \pi^2 f_{\pi}^2 \approx 0.67~GeV^2$
in this monopole formula is
close to $m_{\rho}^2$. 
Recent experimental data \cite{CLEOnew} from CLEO 
are below the BL-curve and 
are by almost a factor of 2 lower than the 
value for the CZ wave function.
This  result apparently excludes  
the CZ DA and suggests that the pion DA 
may be  even narrower than $\varphi_{\pi}^{\rm as}(x)$.  
Since the CZ model is often perceived  
as  a direct consequence from  
QCD sum rules, 
 the experimental evidence in favor
of a narrow DA may be treated as a failure of 
the QCD sum rule approach. 
One should remember, however, that 
accuracy of QCD sum rules strongly
depends on the specific hadronic characteristics 
to which the sum rule technique is applied.  
Long ago,  in papers \cite{mr} written with S. Mikhailov,
we argued that CZ sum rules are very unreliable,
with the results strongly depending on the assumptions
about the size of higher terms in 
the operator product expansion (OPE).

\section{QCD sum rules}

QCD  sum rules   \cite{svz} are based on 
quark-hadron  duality,
$i.e.,$  possibility to describe the same 
object  in terms of either quark or hadronic fields.
To calculate 
$f_{\pi}$, consider  the $p_{\mu}p_{\nu}$-part  of the 
correlator of two axial currents:
\be 
\Pi^{\mu\nu}(p) =
i \int e^{ipx}  \langle 0 | T \,(j_{5\mu}^+(x)\,j_{5\nu}^-(0)
\,)|\,0\rangle\, d^4 x = 
p_{\mu}p_{\nu}\Pi_2(p^2)-g_{\mu\nu}\Pi_1(p^2).
\label{eq:Pi} 
\end{equation}
The dispersion relation represents $\Pi_2(p^2)$ as an integral  over hadronic spectrum 
\be
 \Pi_2(p^2)= \frac1{\pi}\int_0^{\infty}\frac{ \rho^{\rm hadron}(s)}{s-p^2}ds + 
{\rm ``subtractions"} 
\label{eq:PiDR}
\end{equation}
with the spectral
density $\rho^{\rm hadron}(s)$  determined by projections 
of the axial current onto
hadronic states
($
\langle 0 | j_{5\mu} (0) |\pi;  P \rangle = i f_{\pi} P_{\mu},
$ 
$etc.$):
\be
\rho^{\rm hadron}(s) = \pi f_{\pi}^2 \delta(s-m_{\pi}^2) + \pi f_{A_1}^2
\delta(s-m_{A_1}^2)  + ``{\rm higher \   states}" 
\label{eq:rhohadron}
\end{equation}
($f_{\pi}^{\exp} \approx 130 \, MeV$ in our normalization).
On the other hand, when the probing virtuality  is negative and large,
one can use the OPE
\be
\Pi_2(p^2) = \Pi_2^{\rm pert}(p^2) + \frac{A}{p^4} \langle \alpha_s GG \rangle 
+ \frac{B}{p^6} \alpha_s \langle \bar qq \rangle^2  + \ldots
\label{eq:PiOPE}
\end{equation}
where $\Pi_2^{\rm pert}(p^2) \equiv \Pi_2^{\rm quark} (p^2)$ is the 
perturbative  version  of   $\Pi_2 (p^2)$ given by a sum of 
PQCD Feynman diagrams while the condensate terms 
$\langle GG \rangle$,  $\langle \bar qq \rangle$, $etc.,$ 
(with  calculable coefficients
$A, B, etc.$ )  
describe/parameterize  the  nontrivial structure of the QCD vacuum.
The quark amplitude $\Pi_2^{\rm quark}(p^2)$,  can also be written in 
the  dispersion representation   (\ref{eq:PiDR}),  with $\rho(s)$ 
substituted by its perturbative analogue 
$
\rho^{\rm quark}(s)= \frac1{4 \pi} \left ( 1 + \frac{\alpha_s}{\pi} 
+ \ldots \right ) 
$
(quark masses neglected). Hence,  the condensate 
terms describe  the difference between  the 
quark and hadron spectra. 
Treating the condensate values as  known,  
one can try to construct  a model for the hadronic spectrum.
The simplest model 
is to approximate all the higher resonances including the  $A_1$ 
 by the quark  spectral density  starting at some 
effective threshold $s_0$:  
\be
\rho^{\rm hadron}(s) \approx \pi f_{\pi}^2 
\delta(s-m_{\pi}^2) + \rho^{\rm quark}(s) \, 
 \theta(s \geq s_0) .
\label{eq:rhomodelPi} 
\end{equation}
Neglecting the pion mass and 
using the standard values for the condensates
$\langle GG \rangle$, $\langle \bar qq \rangle^2$,
one should adjust  $s_0$ to get an (almost) constant  result
for the rhs of the  SVZ-borelized version
of the sum rule 
\begin{equation}
{f_{\pi}^2} = \frac1{\pi} \int_0^{s_0} \rho^{\rm quark}(s) e^{-s/M^2} {ds}
 \  +\frac{\alpha_s\langle GG\rangle}{12\pi M^2}
		  +\frac{176}{81}\frac{\pi\alpha_s\langle\bar qq\rangle^2}{M^4}
		    + \ldots  \,   .
\label{eq:fpisumruleborel}
\end{equation}
The magnitude  of   $f_{\pi}$ extracted in this way,  is very close
to its  experimental value $f_{\pi}^{\rm exp} \approx 130 \, MeV.$
Changing the  values of the condensates, 
one would get the best $M^2$-stability for a different 
$s_0$, and the resulting value of  $f_{\pi}$ would also change.
Correlation  
between the fitted values of $f_{\pi}$ and $s_0$ 
is manifest in the 
$M^2 \to \infty$ limit of the sum rule  
\be
f_{\pi}^2 = \frac1{\pi} \int_0^{s_0} \rho^{\rm quark}(s) \, ds,
\label{eq:fpiLD}
\end{equation}
giving  a local duality relation which 
 states that two densities $\rho^{\rm quark}(s)$ and $\rho^{\rm hadron}(s)$
give the same result if one integrates 
them over the  {\it appropriate} duality interval $s_0$.
The role of the condensates was to determine the size of the duality
interval $s_0$, but after it was fixed,
one can write down the relation (\ref{eq:fpiLD}) which does not 
involve the condensates.
 In the  lowest order, 
$\rho^{\rm quark}_0(s) = 1/4\pi$, which gives 
$
s_0 = 4\pi^2 f_{\pi}^2 \, .
$
Note, that this is 
exactly the combination which appeared 
in the Brodsky-Lepage interpolation formula.

\section{CZ sum rules and pion DA}

 Chernyak  and 
A. Zhitnitsky  \cite{cz82} proposed 
to use QCD sum rules for calculating 
next moments $\langle\xi^N\rangle$ 
  (where $\xi =2x-1$) of the pion DA. They extracted  
$\langle\xi^2\rangle$ and $\langle\xi^4\rangle$ 
from the relevant SR  
\be
f_\pi^2\langle\xi^N\rangle=\frac{3 M^2}{4\pi^2}
\frac{(1-e^{-s_0/M^2})}{(N+1)(N+3)}
		  +\frac{\alpha_s\langle GG\rangle}{12\pi M^2}
		  +\frac{16}{81}\frac{\pi\alpha_s\langle\bar qq\rangle^2}{M^4}
		    (11+4N)      \label{eq:czsr}
\ee
 precisely in  the 
same way as the $f_{\pi}$ value. 
Note that the scale determining the 
magnitude of the hadronic  parameters  
  is   
settled by the ratios of the condensate contributions  to the  
perturbative term. If the condensate contributions in the CZ  sum  rule 
(\ref{eq:czsr}) would have the same $N$-behavior as the perturbative term,  
then the $N$-dependence of  $\langle\xi^N\rangle$  
would be determined by the  overall 
factor $3/(N+1)(N+3)$     
and the resulting wave function 
$\varphi^{\rm as}_\pi(x)=6 f_\pi x(1-x)$ would coincide 
with the asymptotic form.
However,  the  ratios  of 
the $\langle\bar qq\rangle$  and $\langle GG\rangle$-corrections 
to the  perturbative  
term  in Eq. (\ref{eq:czsr}) are 
growing functions of $N$. In particular, in the $\langle\bar qq\rangle$
case, the above mentioned ratio for $N=2$   is  by factor  
95/11 larger than that in the $N=0$ case. For $N=4$ the
enhancement factor equals 315/11. As a result, 
the effective vacuum scales of $({\em mass})^2$    
dimension are by factors $(95/11)^{1/3}\approx 2.1$ and 
$(315/11)^{1/3}\approx 3.1$ larger than  that 
for the $N=0$ case. Approximately the same factors ($5^{1/2} \approx 2.2$
 and $(35/3)^{1/2} \approx 3.4$) result   from 
the gluon condensate term. Hence, the  parameters $s_0^{(N)}$  
and the combinations $f_\pi^2\langle\xi^N\rangle$ 
straightforwardly extracted from the SR (\ref{eq:czsr}) 
are enhanced compared to 
 $s_0^{N=0}\approx 0.7\,GeV^2$ and 
$3f_{\pi}^2/(N+1)(N+3)$, resp.,  
   by the factors  2 \,(for $N=2$) 
and 3 \,(for $N=4$). These are just the results given in Ref. \cite{cz82}. 
To clarify the assumptions implied by such a procedure,
we rewrite the CZ sum rule  using the standard numerical 
values for the condensates:
\bea
&& \int_0^{\infty} \rho_N (s) e^{-s/M^2} ds = \frac{M^2}{4\pi^2}  \left [\frac{3}{(N+1)(N+3)} 
		\right. \nonumber \\ && \hspace{1cm} \left. 
  + \, 0.1 \left( \frac{0.6}{ M^2} \right )^2  
		  + 0.22 \left (1+ \frac{4N}{11} \right ) 
                 \left( \frac{0.6}{ M^2} \right )^3  \right ] \, .
		          \label{eq:czsrnum}
\eea
Taking first $N=0$, we see that  for $M^2 =0.6$ GeV$^2$ the condensate corrections are
by factor 3 smaller than the perturbative term while the exponential $e^{-s/M^2}$  
suppresses the $A_1$ contribution by  factor 14 compared  to the pion one. 
Hence, the sum rule looks very reliable since 
power corrections are small in the region where 
the $s$-integral  is dominated by the pion. 
Taking the ``first resonance plus effective continuum'' model for the spectrum
and fitting the sum rule in the $M^2 > 0.6$ GeV$^2$ region 
gives $s_0 \approx   0.75$ GeV$^2$ for the effective threshold,
i.e. at the threshold 
 the exponential $e^{-s_0/M^2}$ provides   $1/3$ suppression factor for $M^2 =0.6$ GeV$^2$,
which ensures that the result for $f_{\pi}^2$  is not very sensitive 
to the model chosen for the higher states.

Now,  taking $N=2$, we observe that for $M^2=0.6$ GeV$^2$ 
the condensate corrections are
by factor 2.4 {\it larger}  than the perturbative term:
the $1/M^2$ expansion  is apparently useless at such a 
value of $M^2$. To bring the size of condensate corrections
to less than 1/3 of the perturbative term, one should take $M^2 > 1.2$ GeV$^2$.
However, for  such large $M^2$ values the  exponential $e^{-s/M^2}$  
gives practically no suppression at the ``old'' effective threshold,
and results for $\langle\xi^2\rangle$  would strongly depend on the
model for higher states.  In particular, the
  ``first  resonance plus effective continuum'' ansatz gives  
\mbox{$s_0^{(2)} \approx 1.5 $ GeV$^2$} and $\langle\xi^2\rangle \approx 0.4$
which means that with respect to $\langle\xi^2\rangle$ 
the pion is dual to much wider interval $0 <s <  1.5 $ GeV$^2$. 
 For $N=4$ the duality interval obtained in this way 
is even wider: $s_0^{(4)} \approx 2.2 $ GeV$^2$, i.e.,
the effective continuum threshold is assumed to be well above the $A_1$ location. 

Of course, one cannot exclude a priori that 
a different correlator has a  different  shape of spectral density. 
Ideally, having the full  expression for the right-hand side 
of the sum rule   one could find out   $\rho_N (s)$ exactly.
Having just few terms of the $1/M^2$-expansion,
we can only construct an approximation for the spectrum,
the precision of which depends on the relative magnitude 
of the neglected higher terms. The CZ-procedure is equivalent 
to assumption that two condensate terms included in their sum 
rule dominate the expansion  not only for $N=0$ but also for $N=2$.
In fact, it  is  impossible to check 
by a direct calculation whether this assumption is true or not,
because the number of possible condensates  explodes when 
their dimension increases, and there is no reliable 
way to determine their values.  Still, it is rather 
easy to establish that coefficients accompanying 
the  condensates $\langle\bar q D^2q \, \bar q q\rangle $ 
with two covariant derivatives $D$ behave 
like  $N^3$ for large $N$, i.e. have even larger $N$-dependent
enhancement compared to the perturbative term. In general,
the coefficients for $\langle\bar q (D^2)^n  q \, \bar q q\rangle $  
condensates behave 
like  $N^{n+1}$ for large $N$. Hence,  one would rather
expect that  there are {\it large} higher-condensate corrections
   to the $\langle \xi^2 \rangle $ sum rule.  
Only some miraculous cancellation 
can make them small.  No  reason for such a cancellation was given.

To summarize: if one takes the CZ sum rule at face value,
i.e., assumes that there are no essential corrections to it, the 
fitting procedure  would produce the large CZ value for $\langle \xi^2 \rangle $. 
However, since the perturbative term decreases with  $N$
while the condensate terms rapidly increase with $N$,  the CZ sum rules
for $N \leq 2 $ 
is an obvious  case when one {\it must} expect essential corrections. 

\section{Nonlocal condensates}

It is also instructive  to write  the  SR   for  the  pion 
DA $\varphi_{\pi}(x)$ itself  \cite{mr}:
\bea
&& f_\pi^2\varphi_\pi(x)=\frac{M^2}{4\pi^2}(1-e^{-s_0/M^2})\varphi_{\pi}^{\rm as}(x)
  +\frac{\alpha_s\langle GG\rangle}{24\pi M^2}[\delta(x)+\delta(1-x)]
\nonumber \\
		&& \hspace{1cm}  +\frac{8}{81}\frac{\pi\alpha_s\langle\bar
		   qq\rangle^2}{M^4}
\{11[\delta(x)+\delta(1-x)]+2[\delta^\prime(x)+\delta^\prime(1-x)]\}. 
                         \label{eq:wfsr}
\eea  
The $O(1)$ and $O(N)$ terms in Eq.~(\ref{eq:czsr}) correspond to the 
$\delta(x)$ and $\delta^\prime(x)$-terms  in 
Eq.(\ref{eq:wfsr})
indicating that the vacuum fields are 
carrying zero fraction of the pion momentum. 
The  
operator product expansion (underlying eqs.(\ref{eq:czsr}),(\ref{eq:wfsr})) 
is, in fact, a power 
series expansion over small momenta $k$ of vacuum quarks  and  
gluons. Retaining only the $\langle\bar qq\rangle$  
and $\langle GG\rangle$-terms (like in  eqs.(\ref{eq:czsr}),(\ref{eq:wfsr})) 
is just equivalent to the assumption that $k$ is not  simply  small 
but exactly equals zero. 
However, it is much more reasonable to expect that the  vacuum 
quanta have a smooth distribution  with a finite width $\mu$. 
In configuration space, this means that vacuum fluctuations have  a  
finite correlation length of the order of $1/\mu$, so  that  the  
two-point condensates like $\langle\bar q(0)q(z)\rangle$ 
die away for $|z|$  large compared 
to $1/\mu$.  In the OPE,    $\langle\bar q(0)q(z)\rangle$ is 
expanded in powers of $z$ and   
the first   term  $\langle\bar q(0)q(0)\rangle$ 
produces eventually  the 
$\delta(x)$-term, while higher  $\langle\bar q(0)(D^2)^n q(0)\rangle$ 
terms  give $\delta^n(x)$ contributions resulting 
in $N^n$ factors in the $\langle \xi^N \rangle$ sum  rule.
In other words, arranging the $1/M^2$  expansion 
through the OPE in terms of local operators,
one automatically  obtains  $\langle \xi^N \rangle$ 
in the form of Taylor expansion in $N$.  
Even if the condensate contribution to the $\langle \xi^N \rangle$ 
sum rule is a rapidly decreasing 
function of $N$ (which must be the case for any smooth
function of $\xi$), the OPE gives it as   a Taylor series 
in   $N^n$ whose terms rapidly increase with $N$. 
In such a situation, it is obviously risky 
to take just the first term of the expansion,
e.g., the quark condensate $(11+4N)$ factor 
 may well be just the first term of something like
$(11+4N) \exp[-N \lambda^2 /M^2]$ with  much smaller value 
for $N=2$ than one would expect from $(11+4N)$.
How much smaller, depends on the value of the scale $\lambda^2$. 
The size of the correlation length of vacuum fluctuations 
can be estimated using  the standard 
value   \cite{belioffe}  
$
\lambda_q^2 \equiv \langle\bar q D^2q\rangle / 
\langle\bar qq\rangle = 0.4 \pm 0.1 \, GeV^2   
$
for the average virtuality of  the  vacuum  quarks. 
One can see that it is  
not   small   compared   to   the   relevant  hadronic   scale 
\mbox{$s_0^{N=0} \approx  4\pi^2f_{\pi}^2 = 0.7 \, GeV^2 $, }
and  the                       
constant-field approximation for  the  vacuum  fields  is 
not safe. 
 Using the  exponential model 
\mbox{$\langle\bar q(0)q(z)\rangle= \langle\bar qq\rangle \exp [z^2 \lambda_q^2/2]$}
for the  nonlocal  condensate gives  a QCD sum rule  producing the wave functions 
very close to the asymptotic \mbox{ones\cite{mr}.}  
This study suggests that  the humpy form of 
the CZ wave function is a mere consequence of the approximation
that vacuum quarks have zero momentum. 

\section{QCD sum rule for the $\gamma^* \gamma \to \pi^o$ form factor}

Another evidence that the pion DA is close to 
its asymptotic shape is given by a direct QCD sum rule
analysis \cite{pl,rr} of the $\gamma^* \gamma \to \pi^o$ transition form factor.
In this case, one should consider the 
three-point correlation function
\begin{equation}
{\cal F}_{\alpha\mu \nu}(q_1,q_2)= 2 \pi i
\int
\langle 0 |T\left\{j_{\alpha}^5(Y) J_{\mu }(X)\,J_{\nu}(0)\right\}| 0 \rangle
e^{-iq_{1}X}\,e^{ipY}  d^4X\,d^4Y \,  ,
\label{eq:corr}
\end{equation}
where $J_{\mu}$ is the electromagnetic current.  
The operator product expansion is simpler when both 
photon virtualities $q^2$  are  large:
$q^2,Q^2 \ge 1~GeV^2$.
QCD sum rule in this kinematics is given by 
\begin{eqnarray}
\pi f_{\pi} \mbox{$F_{\gamma^*\gamma^*\pi^\circ}$}(q^2,Q^2)=
 2\int_0^{s_o} ds \, e^{-s/{M^2}}
\int_0^1 \frac{x\bar{x}(xQ^2+ \bar x q^2)^2}
{[s{x}\bar{x}+xQ^2+ \bar x q^2]^3} \,dx  \,
\nonumber \\
+\frac{\pi^2}{9}
{\langle \frac{\alpha_s}{\pi}GG \rangle}
\left(\frac{1}{2M^2 Q^2} + \frac{1}{2M^2 q^2}
 - \frac1{Q^2 q^2}\right)
 \nonumber\\
+ \frac{64}{243}\pi^3\alpha_s{\langle \bar{q}q\rangle}^2
\left( \frac1{M^4}
\left [ \frac{Q^2}{q^4}+ \frac9{2q^2}+\frac9{2Q^2}+\frac{q^2}{Q^4} \right ] +
\frac9{Q^2 q^4} +\frac9{Q^4 q^2} \right )  .
\label{eq:SRLarge}
\end{eqnarray}
 Keeping only the leading  $O(1/Q^2$ and $ 1/q^2)$-terms 
one can rewrite it as 
\begin{eqnarray}
&&F_{\gamma^*\gamma^*\pi^\circ}^{LO}(q^2,Q^2)
=\frac{4\pi}{3f_{\pi}}
   \int_0^1 \frac{dx}{ ( xQ^2 + \bar x q^2)} \,
\left \{ \frac{3M^2}{2\pi^2}(1-e^{-s_0/M^2}) x\bar{x}
\right. \nonumber\\
&&+ \frac{1}{24M^2}
\langle \frac{\alpha_s}{\pi}GG\rangle [\delta(x) + \delta (\bar{x})]
\nonumber\\
&&+\left. \frac{8}{81M^4}\pi\alpha_s{\langle \bar{q}q\rangle}^2
 \biggl ( 11[\delta(x) + \delta (\bar{x})] +
2[\delta^{\prime}(x) + \delta ^{\prime}(\bar{x})]
\biggr ) \right \}
\label{eq:SRlargeQ2wf}.
\end{eqnarray}
Note, that the expression in  curly brackets
coincides with the QCD sum rule (\ref{eq:wfsr}) for
the pion DA
$f_{\pi} \varphi_{\pi}(x)$.  
Hence,
the QCD sum rule approach exactly 
reproduces  the PQCD result (\ref{eq:gg*pipqcd}).
One may be tempted   to  get a 
QCD sum rule for the integral $I$ by taking  $q^2=0$
in Eq.(\ref{eq:SRLarge}). The attempt  is ruined by
 power singularities
$1/q^2$, $1/q^4$
in the condensate terms.
Moreover, the perturbative term in the small-$q^2$ region has
logarithms $\log q^2$
which  are a typical example
of mass singularities (see, $e.g.,$  \cite{georgietal}).  
All these infrared sensitive terms are produced 
in a regime when the hard momentum flow
bypasses the soft photon vertex, $i.e.$,
the EM current $J_{\mu }(X)$ of
 the low-virtuality photon is
far away from the two other  currents $J(0),j^5(Y)$. 
It is also important to observe that power singularities $1/q^2$, $1/q^4$
 are generated  precisely   by  the  same
$\delta(x)$ and $\delta'(x)$ terms in Eq.(\ref{eq:SRlargeQ2wf})
which  generate the two-hump form
for $\varphi_{\pi}(x)$ in the CZ-approach  \cite{cz82}.
As shown in Ref.\cite{mr}, the humps disappear if 
one treats the $\delta(x)$ and $\delta'(x)$ terms
as the first terms of a formal expansion $\Phi (x) \sim 
\sum a_n \delta^n (x)$  of  smooth
functions $\Phi (x)$.  
Similarly, the $1/q^2$ singularity   
can be understood as the first term of the large-$q^2$
expansion of a term like $1/(q^2+m_{\rho}^2)$
in powers of $1/q^2$.  
However,  constructing 
$\Phi (x)$ from  two first terms of such expansion
is a strongly  model-dependent procedure.
On the other hand,   the   small-$q^2$ 
behavior of the three-point function is
rather severely constrained by known structure 
of the  physical spectrum in the EM-current channel.
The procedure developed  in Refs.\cite{pl,rr}  allows 
 to subtract all the small-$q^2$  singularities from
the coefficient functions of the original OPE for the 3-point
correlation function Eq.(\ref{eq:corr}). They are absorbed in this
approach by  universal bilocal correlators, 
which can be also interpreted
as moments of the DAs  for  (almost) real photon
$$
\int_0^1 y^n \phi_{\gamma}^{(i)} (y,q^2) \sim
\Pi^{(i)}_n(q^2) = \int e^{iq_1 X} \langle 0| T \{ J_{\mu }(X)
{\cal O}_n^{(i)}  (0) \}| 0 \rangle d^4X ,
$$
where ${\cal O}_n^{(i)} (0)$ are operators of leading and next-to-leading
twist with $n$ covariant derivatives  \cite{pl,rr}.
The bilocal contribution to the 3-point function Eq.(\ref{eq:corr})
can be written in a ``parton'' form as a convolution of the 
photon DAs and some coefficient functions. The latter originate from
a light cone OPE for the product  $T\{ J(0) j^5(Y) \}$.
The
amplitude ${\cal F}$   is now  a  sum of its
purely  short-distance ($SD$) (regular for $q^2=0$) and bilocal ($B$) parts.
Getting the $q^2 \to 0$ limit of $\Pi_n^{(i)}(q_1)$
requires a nonperturbative input obtained from an auxiliary QCD sum rule.
After all the modifications outlined above are made, one can write 
the QCD sum rule for the $\gamma \gamma^* \to \pi^0$ form factor
in the $q^2=0$ limit:
\begin{eqnarray}
&& \pi f_{\pi} F_{\gamma \gamma^* \pi^0}(Q^2) =
\int_0^{s_0}
\left \{ 
1 - 2 \frac{Q^2-2s}{(s+Q^2)^2}
\left (s_{\rho} - \frac{s_{\rho}^2}{2 m_{\rho}^2} \right )
\right.  \nonumber \\ && +  \left.
  2\frac{Q^4-6sQ^2+3s^2}{(s+Q^2)^4} \left (\frac{s_{\rho}^2}{2}
 - \frac{s_{\rho}^3}{3  m_{\rho}^2} \right )
\right \} 
 e^{-s/M^2}
\frac{Q^2 ds }{(s+Q^2)^2}
 \nonumber \\
&& + \frac{\pi^2}{9}
{\langle \frac{\alpha_s}{\pi}GG \rangle}
\left \{ 
\frac{1}{2 Q^2 M^2} + \frac{1}{Q^4}
- 2 \int_0^{s_0} e^{-s/M^2} \frac{ds }{(s+Q^2)^3}
\right \} 
\label{eq:finsr} \\ && +
 \frac{64}{27}\pi^3\alpha_s{\langle \bar{q}q\rangle}^2
\lim_{\lambda^2 \to 0}
\left \{ 
\frac1{2Q^2 M^4}
+ \frac{12}{Q^4 m_{\rho}^2 }
\left [ 
\log \frac{Q^2}{\lambda ^2} -2
\right.  \right. \nonumber \\ && + \left.  \left.
 \int_0^{s_0} e^{-s/M^2}
\left ( 
\frac{s^2+3sQ^2+4Q^4} {(s+Q^2)^3} - \frac1{s+\lambda ^2}
\right) ds 
\right] 
\right. 
\nonumber \\
&& - 
\left.  
\frac4{Q^6}
\left [ 
\log \frac{Q^2}{\lambda^2} -3+
\int_0^{s_0} e^{-s/M^2}
\left (  
\frac{s^2+3sQ^2+6Q^4} {(s+Q^2)^3} - \frac1{s+\lambda ^2}
\right) ds 
\right] 
\right \} .
 \nonumber
\end{eqnarray}
Here the bilocal contributions are modeled by  asymptotic form for
the $\rho$-meson DAs. They are  approximately
dual to the corresponding  perturbative contribution with 
the $\rho$-meson duality interval \cite{svz}
 $s_{\rho}=1.5~GeV^2$.
\begin{figure}[htb]
\mbox{
   \epsfxsize=8.0cm
 \epsffile{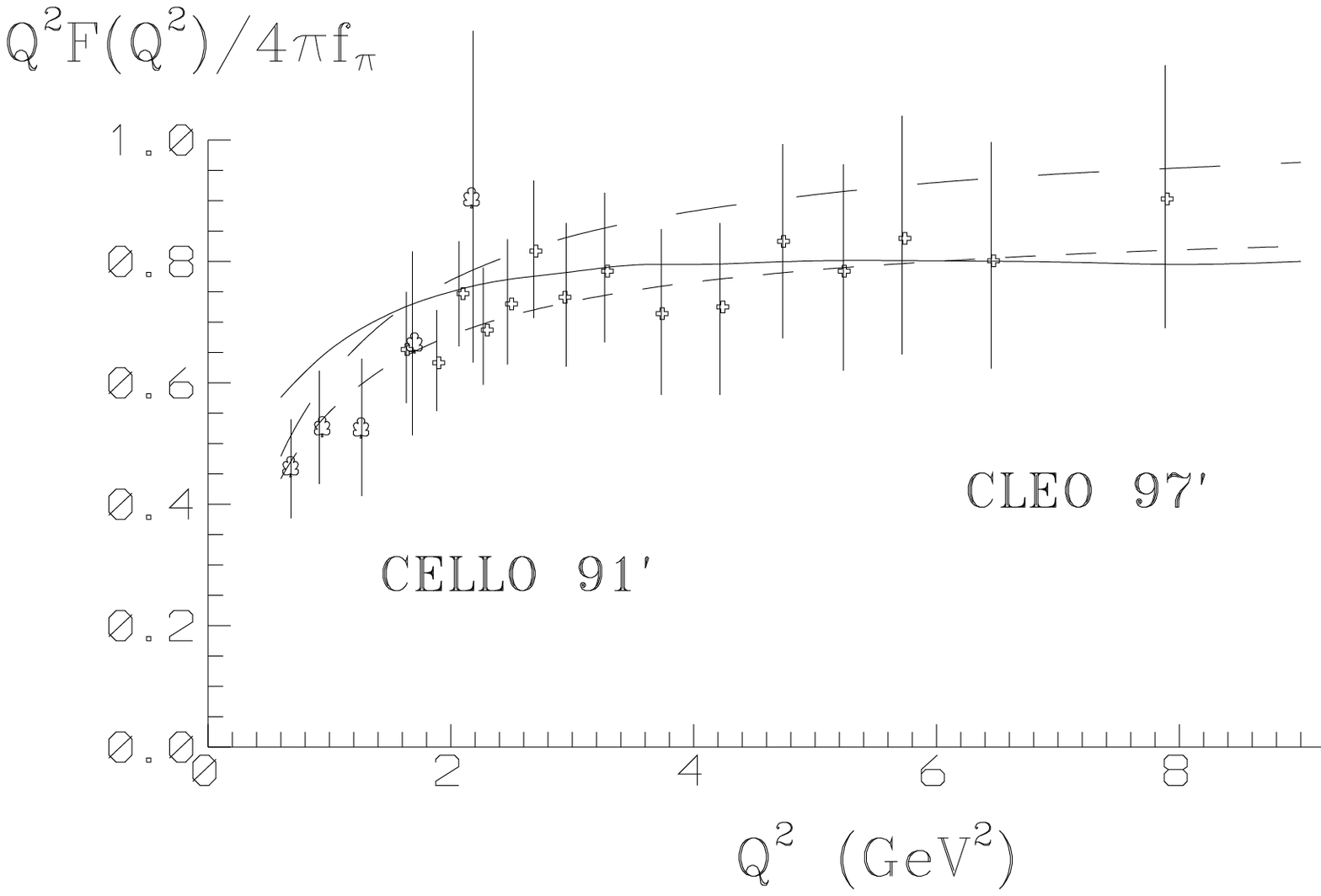}  }
  \vspace{-5.8cm}
{\caption{\label{fig:3} }}
\end{figure}
The results of fitting procedure for 
(\ref{eq:finsr}) favor  the value
$s_0 \approx 0.7$   GeV$^2$ as  the effective threshold  \cite{pl,rr}. 
For this reason, the results of  our calculations 
are well approximated by 
the local duality prescription  \cite{radacta95}
\be
\pi f_{\pi} F^{\rm LD}_{\gamma \gamma^* \pi^0}(Q^2) =
\int_0^{s_0} \rho^{\rm quark}(s,Q^2) ds = \frac1{1+Q^2/s_0} \, 
\ee
which coincides for $s_0 =4 \pi^2 f_{\pi}^2$ with the BL interpolation formula.
In Fig.\ref{fig:3}, 
we present our  curve (solid line) for
$Q^2F_{\gamma \gamma^* \pi^0}(Q^2)/4\pi f_{\pi}$
calculated from Eq.(\ref{eq:finsr}) for $s_0 = 0.7 \, GeV^2$ 
and $M^2 = 0.8\, GeV^2$. 
One can observe very good agreement with
the new CLEO data  \cite{CLEOnew}.
It is also rather close to the
BL interpolation/local duality 
formula  (long-dashed line)
and  the $\rho$-pole  approximation (short-dashed line)
$\pi f_{\pi} F^{VMD}(Q^2) = 1/(1+Q^2/m_{\rho}^2)$.
It should be noted that
the  $Q^2$-dependence of the $\rho$-pole type  emerges
due to the fact that the pion
duality interval $s_0 \approx 0.67 \, GeV^2$
is numerically  close to $m_{\rho}^2\approx 0.6\,GeV^2$.
In the region  $Q^2 > Q^2_{*} \sim 3 \, GeV^2$, our curve for
$Q^2F_{\gamma \gamma^* \pi^0}(Q^2)$
 is practically  constant, supporting
 the PQCD expectation (\ref{eq:gg*pipqcd}).  The
absolute magnitude  of our prediction  gives
  $I \approx 2.4$ for the $I$-integral with an
accuracy of about $20\%$.
Comparing the value  $I=2.4$ with
 $I^{\rm as}=3$ and $I^{\rm CZ}=5$, we
conclude  that our result favours
a  pion  DA
which is narrower than the asymptotic form.
Parametrizing the width of $\varphi_{\pi}(x)$  by
 a simple model \mbox{$\varphi_{\pi}(x) \sim [x(1-x)]^n$,}
we obtain  that  $I=2.4$
corresponds to $n=2.5$.
The second moment  
\mbox{$ \langle \xi^2\rangle \equiv \langle (x - \bar x)^ 2\rangle$}
for such a function  
is 0.125
(recall that $\langle \xi^2\rangle^{\rm as}=0.2$
while $\langle \xi^2\rangle^{\rm CZ}=0.43$)  which agrees  with
the lattice calculation  \cite{lattice}.

Thus, the old claim \cite{mr} that the CZ sum rules \cite{cz82} 
for the moments of DAs are  unreliable
is now  supported both by a direct QCD sum rule 
calculation  of the $\gamma^* \gamma \pi^0$
form factor \cite{pl,rr} producing the result corresponding
to a narrow pion DA, and by experimental 
measurement of this form factor \cite{CLEOnew} which 
also favors a pion DA close to the asymptotic form.
Since the  humpy form of the CZ models for the nucleon 
DA's \cite{cz82}  has the same origin 
as in the pion case,  there is  no doubt  that 
the nucleon DA's are also close to the asymptotic ones.
This means that PQCD contributions to nucleon elastic and transition
form factors are tiny at available and reachable energies.

I thank R. Ruskov and I. Musatov for collaboration and  V.Savinov
for correspondence.
This  work   was supported
by the US Department of Energy under contract DE-AC05-84ER40150.

\end{document}